\documentclass[showpacs,preprintnumbers,superscriptaddress,nofootinbib,prl]{revtex4}
\usepackage{amsmath}
\usepackage{portland}
\usepackage[dvips]{graphicx}
\usepackage{dcolumn}
\usepackage{bm}
\usepackage{amsfonts}
\usepackage{rotating}
\usepackage{amssymb}
\usepackage{scalefnt}
\usepackage[center]{caption2}

\input{tcilatex}

\begin{document}

\preprint{}
\title{Improved Exponential model with pairing attenuation and the
bakbending phenomenon}
\author{H. A. Alhendi}
\email{alhendi@ksu.edu.sa}
\affiliation{Department of Physics and Astronomy, College of Science, King Saud
University, P.O. Box 2455, Riyadh 11454, Saudi Arabia}
\author{H. H. Alharbi}
\email{alharbi@kacst.edu.sa}
\affiliation{National Center for Mathematics and Physics, KACST, P.O. Box 6086, Riyadh
11442, Saudi Arabia,}
\author{S. U. El-Kameesy}
\email{kameesey@ksu.edu.sa}
\affiliation{Department of Physics and Astronomy, College of Science, King Saud
University, P.O. Box 2455, Riyadh 11454, Saudi Arabia}
\date{\today }

\begin{abstract}
{\small A modified version of the exponential model with paring attenuation
is proposed, and used to describe successfully the backbending of the moment
of inertia, in even-even nuclei, not only in well-deformed nuclei but also
in slightly deformed nuclei. The model remarkably fits good \ the
experimental observation with a few parameters.}
\end{abstract}

\maketitle

Several models have been remarkably successful in the description of the low
lying collective states in many light, medium and heavy even-even nuclei. In
the last three decades, a large amount of high-spin states of nuclei has
been accumulated. Among these data, an interesting effect has been observed
in the ground state rotational band (yrast states) of some even-even nuclei.
It is called the backbending which occurs as one plots the moment of inertia
versus the square of the rotational frequency. Many efforts \cite%
{Gelberg,Alonso,Faessler,Stephens1,Long,Hsich,seven} within the frame work
of different models have been attempted to understand the mechanism of the
rapid change of the moment of inertia. Three types of explanations have been
proposed for this change. The essential features of these explanations are:
pairing collapse \cite{Mottelson}, rotation alignment \cite%
{Stephens,Wyss,Bengtsson} and centrifugal stretching \cite{Thieburger}. Most
of these features could be described in terms of band mixing or band
crossing.

The rapid increase of the moment of inertia with the rotational frequency
indicates a major modification of the intrinsic structure at the point where
backbending occurs. The fact that the moment of inertia is almost doubled
and is approaching the value of a rigid rotation suggests that the
transition is associated with a considerable reduction in the pair
correlation \cite{Bohr}. The dependence of the nuclear moments of inertia on
pairing correlations (in an exponential form) has been established
previously \cite{Sood}. In an attempt to provide a theoretical basis for
this relation, Sood and Jain \cite{Sood} found that the underlying physical
process is attributed to a transition from quasiparticles in the presence of
pairing, to independent particles as the pairing disappears. The analogy
between the destruction of the pair correlation by coupling at large orbital
angular momentum $I$ and the destruction of the superconductivity by a
magnetic field led them to draw a relation between the pairing gap and the
angular momentum in the following form:

\begin{equation}
\Delta\left( I\right) =\Delta\left( 0\right) \left( 1-\frac{I}{I_{c}}\right)
^{\frac{1}{\nu}}  \label{Pairing}
\end{equation}
with $\nu=2$, such that the pairing gap $\Delta\left( I\right) $ vanishes at
the angular momentum $I_{c}$ This relation was previously obtained for
nuclear problems by Moretto in his statistical model calculations \cite%
{Mortto}. Accordingly, Sood and Jain gave the following expression \ for
rotational energy levels of ground state band:

\begin{equation}
E\left( I\right) =\frac{\hbar^{2}}{2\varphi_{0}}I\left( I+1\right)
e^{\Delta_{0}\sqrt{1-\frac{I}{I_{c}}}}  \label{energy}
\end{equation}

They applied their approach to describe the backbending phenomena in the
well-deformed nuclei region ($150<A<190$), with $\Delta _{0}$ and $\varphi
_{0}$ as free parameters and $I_{c}$ is selected around $18\hbar $.
Satisfactory results have been obtained below the point where backbending
occurs. Making use of their formula it was not however possible to describe
the forward or down-bending region of the $\varphi -\omega ^{2}$ plot.

Further applications of the exponential model with pairing attenuation have
been recently carried out to the superdeformed nuclei in the $A\sim 190$
region \cite{Shan} and satisfactory results have been obtained for
superdeformed bands.

Old and recent theoretical investigations of backbending mechanism have led
to the use of multiparameter models, such as the various extensions of VMI,
and have yielded qualitatively different results at different mass regions
where the corresponding deformations change considerably from slightly
deformed to superdeformed nuclei.

We present, here, a simple three parameter modified formula based on the
exponential model with paring attenuation for the description of the
backbending phenomena, by letting $\nu$ to be a free parameter. We take this
model as a phenomenological one without reverting to its original
derivation. This model not only reasonably describes the backbending of the
moment of inertia in the well-deformed nuclei but its excellent validity
also extends to include the slightly deformed nuclei at $A\sim100$, and
actinide nuclei where the interplay between single particle and collective
aspects maybe expected at $A\symbol{126}240$ as well.

\begin{sidewaystable}[tbp] \centering%

\begingroup
\scalefont{.8}

\caption
{Experimental and calculated energies (in MeV) of levels of ground state bands of even-even Nuclei}%
\begin{tabular}{ccccccccccccccccc}
\hline\hline
$I$ &  & 2$^{+}$ & 4$^{+}$ & 6$^{+}$ & 8$^{+}$ & 10$^{+}$ & 12$^{+}$ & 14$%
^{+}$ & 16$^{+}$ & 18$^{+}$ & 20$^{+}$ & 22$^{+}$ & 24$^{+}$ & 26$^{+}$ & 28$%
^{+}$ & 30$^{+}$ \\ \hline
$^{100}$Zr & Exp. & 0.21253 & 0.564486 & 1.0514 & 1.6874 & 2.4255 & 3.2721 &
4.2085 &  &  &  &  &  &  &  &  \\
& Calc. & 0.18971 & 0.560775 & 1.0676 & 1.6903 & 2.4217 & 3.2621 & 4.2147 &
&  &  &  &  &  &  &  \\
$^{102}$Zr & Exp. & 0.15178 & 0.47828 & 0.96478 & 1.5949 & 2.3515 & 3.2123 &
&  &  &  &  &  &  &  &  \\
& Calc. & 0.14888 & 0.47777 & 0.96625 & 1.5958 & 2.3498 & 3.21288 &  &  &  &
&  &  &  &  &  \\
$^{110}$Pd & Exp. & 0.3738 & 0.92077 & 1.57404 & 2.296 & 3.131 & 4.03 &  &
&  &  &  &  &  &  &  \\
& Calc. & 0.3636 & 0.92497 & 1.57641 & 2.303 & 3.117 & 4.036 &  &  &  &  &
&  &  &  &  \\
$^{112}$Cd & Exp. & 0.61752 & 1.41558 & 2.16803 & 2.88126 & 3.68426 & 4.58739
&  &  &  &  &  &  &  &  &  \\
& Calc. & 0.62014 & 1.41170 & 2.16540 & 2.89430 & 3.67167 & 4.59132 &  &  &
&  &  &  &  &  &  \\
$^{120}$Ba & Exp. & 0.186 & 0.544 & 1.0402 & 1.645 & 2.336 & 3.083 &  &  &
&  &  &  &  &  &  \\
& Calc. & 0.180 & 0.544 & 1.0435 & 1.647 & 2.332 & 3.08461 &  &  &  &  &  &
&  &  &  \\
$^{122}$Xe & Exp. & 0.33118 & 0.8283 & 1.46657 & 2.2173 & 3.0395 & 3.9192 &
4.9 &  &  &  &  &  &  &  &  \\
& Calc. & 0.30529 & 0.8330 & 1.48217 & 2.2166 & 3.0282 & 3.92054 & 4.90232 &
&  &  &  &  &  &  &  \\
$^{130}$Nd & Exp. & 0.15905 & 0.4855 & 0.93994 & 1.4871 & 2.10041 & 2.7639 &
3.4681 & 4.2116 & 5.0209 & 5.9183 & 6.9092 & 7.9939 &  &  &  \\
& Calc. & 0.16598 & 0.4987 & 0.95023 & 1.4884 & 2.09254 & 2.75099 & 3.45914
& 4.21834 & 5.03534 & 5.92255 & 6.89919 & 7.99469 &  &  &  \\
$^{132}$Nd & Exp. & 0.21262 & 0.6098 & 1.1312 & 1.7104 & 2.3091 & 2.9448 &
3.6301 & 4.369 & 5.1799 & 6.0625 & 7.0063 & 8.0078 &  &  &  \\
& Calc. & 0.21452 & 0.6132 & 1.1196 & 1.6913 & 2.3072 & 2.9597 & 3.64989 &
4.38461 & 5.1749 & 6.03505 & 6.98212 & 8.03554 &  &  &  \\
$^{140}$Ba & Exp. & 0.60235 & 1.13059 & 1.6607 & 2.4689 & 3.3836 &  &  &  &
&  &  &  &  &  &  \\
& Calc. & 0.60172 & 1.12058 & 1.6939 & 2.4329 & 3.3963 &  &  &  &  &  &  &
&  &  &  \\
$^{142}$Gd & Exp. & 0.51537 & 1.20907 & 2.00289 & 2.7589 & 3.4094 & 4.1032 &
4.9019 & 5.8119 &  &  &  &  &  &  &  \\
& Calc. & 0.50906 & 1.24572 & 1.99708 & 2.7152 & 3.4140 & 4.12988 & 4.90834
& 5.79996 &  &  &  &  &  &  &  \\
$^{150}$Nd & Exp. & 0.13521 & 0.38145 & 0.7204 & 1.1297 & 1.599 & 2.119 &
2.6825 &  &  &  &  &  &  &  &  \\
& Calc. & 0.12861 & 0.38139 & 0.7226 & 1.1319 & 1.598 & 2.11612 & 2.68409 &
&  &  &  &  &  &  &  \\
$^{152}$Nd & Exp. & 0.07251 & 0.23662 & 0.48395 & 0.8054 & 1.1953 & 1.6476 &
2.1579 & 2.7224 & 3.3357 &  &  &  &  &  &  \\
& Calc. & 0.07327 & 0.23708 & 0.48344 & 0.8050 & 1.1952 & 1.64803 & 2.15828
& 2.72177 & 3.33589 &  &  &  &  &  &  \\
$^{160}$Dy & Exp. & 0.0867882 & 0.283824 & 0.58118 & 0.96685 & 1.42872 &
1.9515 & 2.515 & 3.0917 & 3.6722 &  &  &  &  &  &  \\
& Calc. & 0.0880124 & 0.285594 & 0.58260 & 0.96775 & 1.42823 & 1.94916 &
2.5127 & 3.09646 & 3.8466 &  &  &  &  &  &  \\
$^{162}$Er & Exp. & 0.10204 & 0.32961 & 0.66664 & 1.09668 & 1.60282 & 2.16511
& 2.74572 & 3.2924 & 3.8466 &  &  &  &  &  &  \\
& Calc. & 0.10354 & 0.33250 & 0.67076 & 1.10082 & 1.60339 & 2.15678 & 2.73581
& 3.31008 & 3.84067 &  &  &  &  &  &  \\
$^{170}$Os & Exp. & 0.2867 & 0.7499 & 1.32542 & 1.9458 & 2.5452 &  &  &  &
&  &  &  &  &  &  \\
& Calc. & 0.2745 & 0.7561 & 1.33077 & 1.9380 & 2.5478 &  &  &  &  &  &  &  &
&  &  \\
$^{172}$Os & Exp. & 0.22777 & 0.60617 & 1.05447 & 1.52495 & 2.02387 & 2.5645
& 3.1994 & 3.8233 & 4.5107 &  &  &  &  &  &  \\
& Calc. & 0.23130 & 0.61103 & 1.05447 & 1.52070 & 2.02311 & 2.57309 & 3.19519
& 3.92075 & 4.78664 &  &  &  &  &  &  \\
$^{180}$Os & Exp. & 0.13211 & 0.40862 & 0.79508 & 1.25744 & 1.76757 & 2.30871
& 2.875 &  &  &  &  &  &  &  &  \\
& Calc. & 0.13409 & 0.41108 & 0.79455 & 1.25465 & 1.76702 & 2.31195 & 2.874
&  &  &  &  &  &  &  &  \\
$^{182}$Os & Exp. & 0.127 & 0.4004 & 0.7938 & 1.2778 & 1.8121 & 2.3463 &
2.8409 & 3.3202 &  &  &  &  &  &  &  \\
& Calc. & 0.130 & 0.4076 & 0.8006 & 1.2764 & 1.8012 & 2.3404 & 2.85736 &
3.31402 &  &  &  &  &  &  &  \\
$^{190}$Os & Exp. & 0.186718 & 0.547854 & 1.05038 & 1.66647 & 2.357 &  &  &
&  &  &  &  &  &  &  \\
& Calc. & 0.182002 & 0.549262 & 1.05351 & 1.66316 & 2.358 &  &  &  &  &  &
&  &  &  &  \\
$^{192}$Os & Exp. & 0.205794 & 0.58028 & 1.08923 & 1.70839 & 2.4188 & 3.211
&  &  &  &  &  &  &  &  &  \\
& Calc. & 0.198926 & 0.58146 & 1.09319 & 1.70809 & 2.4154 & 3.213 &  &  &  &
&  &  &  &  &  \\
$^{230}$Th & Exp. & 0.0532 & 0.1741 & 0.3566 & 0.5941 & 0.8797 & 1.2078 &
1.5729 & 1.9715 & 2.3978 & 2.85 & 3.325 & 3.812 &  &  &  \\
& Calc. & 0.0545 & 0.1759 & 0.3578 & 0.5940 & 0.8788 & 1.20665 & 1.57248 &
1.97147 & 2.39907 & 2.85 & 3.323 & 3.812 &  &  &  \\
$^{232}$Th & Exp. & 0.049369 & 0.16212 & 0.3332 & 0.5569 & 0.827 & 1.1371 &
1.4828 & 1.8586 & 2.2629 & 2.6915 & 3.1442 & 3.6196 & 4.1162 & 4.6318 & 5.162
\\
& Calc. & 0.051554 & 0.16614 & 0.3374 & 0.5595 & 0.827 & 1.1353 & 1.4797 &
1.8564 & 2.2618 & 2.6928 & 3.1467 & 3.6216 & 4.1159 & 4.6291 & 5.163 \\
$^{240}$Pu & Exp. & 0.042824 & 0.14169 & 0.294319 & 0.49752 & 0.7478 & 1.0418
& 1.3756 & 1.7456 & 2.152 & 2.591 & 3.061 & 3.56 & 4.088 &  &  \\
& Calc. & 0.044081 & 0.14412 & 0.296843 & 0.49911 & 0.7480 & 1.0403 & 1.3735
& 1.7448 & 2.152 & 2.592 & 3.062 & 3.56 & 4.087 &  &  \\
$^{242}$Pu & Exp. & 0.04454 & 0.1473 & 0.3064 & 0.5181 & 0.7786 & 1.0844 &
1.4317 & 1.8167 & 2.236 & 2.686 & 3.163 & 3.662 & 4.172 &  &  \\
& Calc. & 0.04536 & 0.1488 & 0.3075 & 0.5183 & 0.7781 & 1.0834 & 1.4308 &
1.8166 & 2.237 & 2.687 & 3.163 & 3.660 & 4.173 &  &  \\ \hline\hline
\end{tabular}%
\endgroup
\end{sidewaystable}
\begin{table}[tbp]
\caption{The fitting parameters of the present model (Eq. 3). The fifth
column gives the root mean squre diviation (Eq. 4), and the last column
gives the ratio R4 (R4=E4+/E2+).}\centering
\par
\begin{tabular}{ccccccc}
\hline\hline
Nucleus & $2\varphi _{0}/\hbar ^{2}$ & $\Delta _{0}$ & $\nu $ & $I_{c}$ & 
rmsd & $E_{4^{+}}/E_{2^{+}}$ \\ \hline
$^{100}$Zr & 57.1918 & 0.7389 & 0.114596 & 80 & 0.0411068 & 2.66 \\ 
$^{102}$Zr & 120.206 & 1.1312 & 0.733982 & 80 & 0.0078426 & 3.15 \\ 
$^{110}$Pd & 48.7823 & 1.4309 & 0.114326 & 64 & 0.0115461 & 2.46 \\ 
$^{112}$Cd & 37.1679 & 1.7962 & 0.462671 & 16 & 0.0031577 & 2.29 \\ 
$^{120}$Ba & 87.8122 & 1.0736 & 0.244153 & 80 & 0.0135763 & 2.92 \\ 
$^{122}$Xe & 55.9335 & 1.2866 & 0.122269 & 80 & 0.0299285 & 2.50 \\ 
$^{130}$Nd & 75.7929 & 0.8509 & 0.575297 & 26 & 0.0153101 & 3.05 \\ 
$^{132}$Nd & 77.2153 & 1.1831 & 0.451699 & 30 & 0.0061678 & 2.87 \\ 
$^{140}$Ba & 34.8359 & 2.2409 & 0.118346 & 30 & 0.0119052 & 1.88 \\ 
$^{142}$Gd & 51.9782 & 1.833 & 0.451109 & 22 & 0.0131265 & 2.35 \\ 
$^{150}$Nd & 106.816 & 0.9608 & 0.170712 & 80 & 0.0185105 & 2.82 \\ 
$^{152}$Nd & 104.880 & 0.2774 & 0.920592 & 20 & 0.0035709 & 3.26 \\ 
$^{160}$Dy & 442.245 & 1.8949 & 6.008963 & 26 & 0.005267 & 3.27 \\ 
$^{162}$Er & 597.374 & 2.3679 & 5.397786 & 26 & 0.0067128 & 3.23 \\ 
$^{170}$Os & 154.772 & 2.1634 & 0.253182 & 80 & 0.0194954 & 2.62 \\ 
$^{172}$Os & 72.0404 & 1.287 & 0.45591 & 20 & 0.0229337 & 2.66 \\ 
$^{180}$Os & 510.846 & 2.5195 & 0.742648 & 80 & 0.0061911 & 3.26 \\ 
$^{182}$Os & 17903.0 & 6.0212 & 7.58905 & 28 & 0.0120262 & 3.15 \\ 
$^{190}$Os & 80.8917 & 1.0064 & 0.22128 & 80 & 0.0114673 & 2.93 \\ 
$^{192}$Os & 63.1590 & 0.8944 & 0.13273 & 80 & 0.0137451 & 2.82 \\ 
$^{230}$Th & 457.362 & 1.4557 & 1.14749 & 80 & 0.0076974 & 3.27 \\ 
$^{232}$Th & 184.332 & 0.4939 & 0.90486 & 32 & 0.0135804 & 3.28 \\ 
$^{240}$Pu & 292.084 & 0.7829 & 1.01142 & 80 & 0.0097799 & 3.31 \\ 
$^{242}$Pu & 79930.8 & 6.4192 & 15.3588 & 56 & 0.0059640 & 3.31 \\ 
\hline\hline
\end{tabular}%
\end{table}

In the present work the energy levels of the ground state band, for
even-even nuclei, are obtained from the following expression: 
\begin{equation}
E\left( I\right) =\frac{\hbar^{2}}{2\varphi_{0}}I\left( I+1\right)
e^{\Delta_{0}\left( 1-\frac{I}{I_{c}}\right) ^{\frac{1}{\nu}}}\text{,}
\label{energy2}
\end{equation}
where $\varphi_{0},\Delta_{0}$ and $\nu$ are the three parameters of the
model which are adjusted to give a least squares fit to the experimental
data for low and high angular momenta. To fit these parameters we have
carried out an exponential model least squares fit to the observed energy
levels up to $I=30$. For better predictive power we scanned for optimum $%
I_{c}$ by keeping $\varphi_{0},\Delta_{0}$ and $\nu$ as free parameters,
where $I_{c}$ corresponds to the minimum value of the root mean square
deviation value $\sigma$ given by:

\begin{equation}
\sigma=\sqrt{\frac{1}{N}\sum_{i=1}^{N}\left( 1-\frac{E_{cal}}{E_{\exp}}%
\right) ^{2}}\text{,}  \label{Sigma}
\end{equation}

In the case where $\sigma $ changes exponentially with no minimum we adopt $%
I_{c}$ as the value where the variation of $I_{c}$ does not affect $\sigma $
significantly.

Next we undertake a comparative study between our calculations and the
experimental data through $\varphi-\omega^{2}$ plots. The moment of inertia $%
\varphi$ and the squared rotatonal frequency $\omega^{2}$ are related to the
spin derivatives of the energy through the relations

\begin{equation}
2\varphi/\hbar^{2}=\left( 4I-2\right) /\Delta E_{\gamma}  \label{BB1}
\end{equation}

\begin{equation}
\left( \hbar\omega\right) ^{2}=\left( I^{2}-I+1\right) /\left[ \Delta
E_{\gamma}/\left( 2I-1\right) \right] ^{2}  \label{BB2}
\end{equation}

where

\begin{equation*}
\Delta E_{\gamma}=E\left( I\right) -E\left( I-2\right)
\end{equation*}

The results of the ground state energy levels up to spin $I=30$ for a set of
representative nuclei\ are presented in Table I, with the corresponding
experimental energy levels.\cite{iaea}, and the calculated parameters are
given in Table II.

In Figs. 1,2, and 3, we present the $\varphi-\omega^{2}$ plots for 24 nuclei
covering the mass region from 100 to 242. The backbending displayed in Figs.
1, 2, and 3 shows an excellent agreement between the model predictions and
experimental data even in the forward or downbending regions.

\begin{figure}[htbp]
\centering
\includegraphics*[clip,width=2.7449in, height=3.9029in]{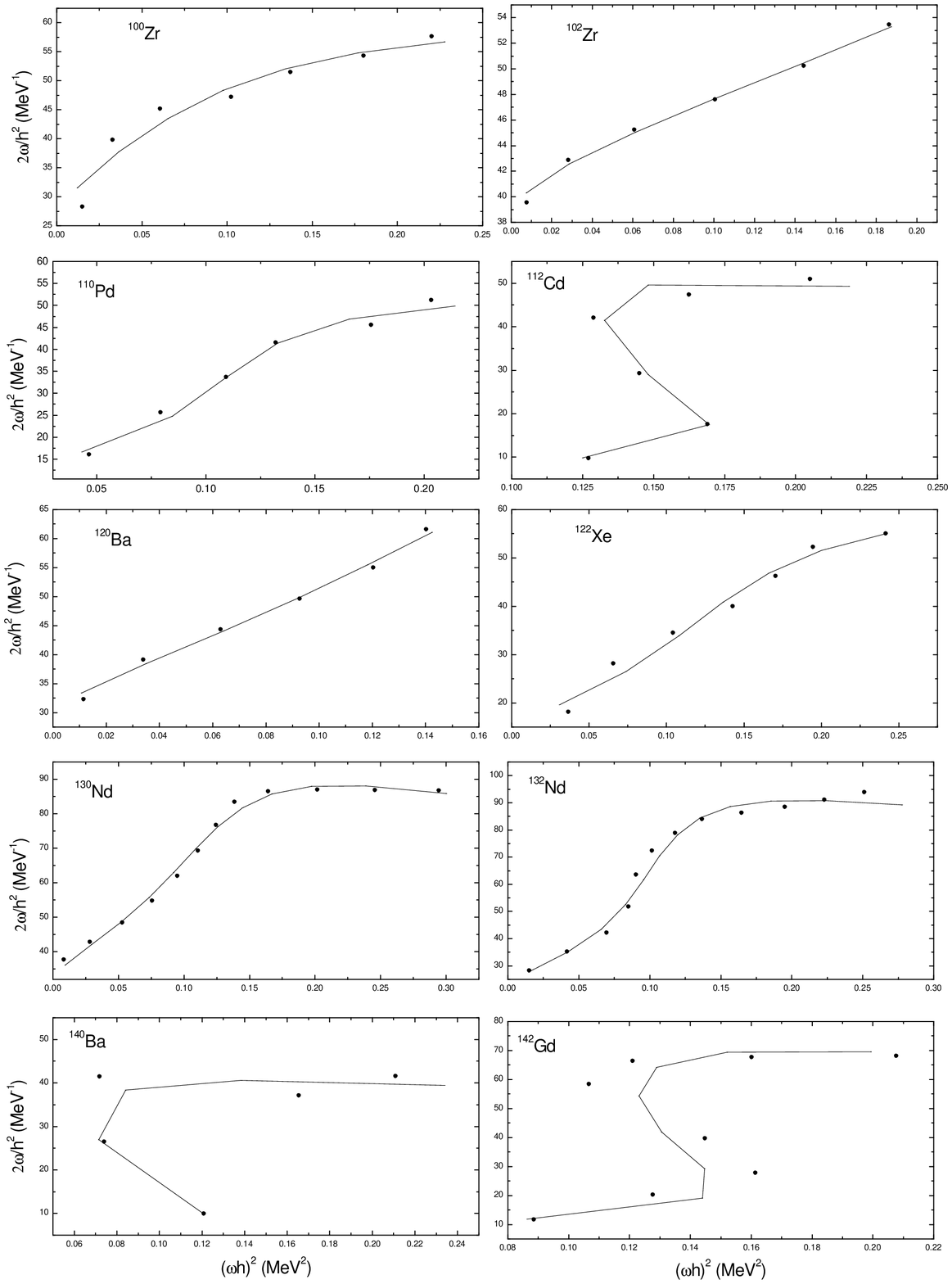}
\caption{The moment of enertia 2$\protect\varphi /\hbar ^{2}$ versus the
square of the rotational frequency ($\hbar \protect\omega )^{2}$ for $%
^{100,102}$Zr, $^{110,112}$Pd, $^{120,140}$ Ba, $^{122}$Xe, $^{130,132}$Nd,
and $^{142}$Cd. Dots represent experimental values}
\label{Fig:Figure1}
\end{figure}

\begin{figure}[htbp]
\centering
\includegraphics*[clip,width=2.7449in, height=3.9029in]{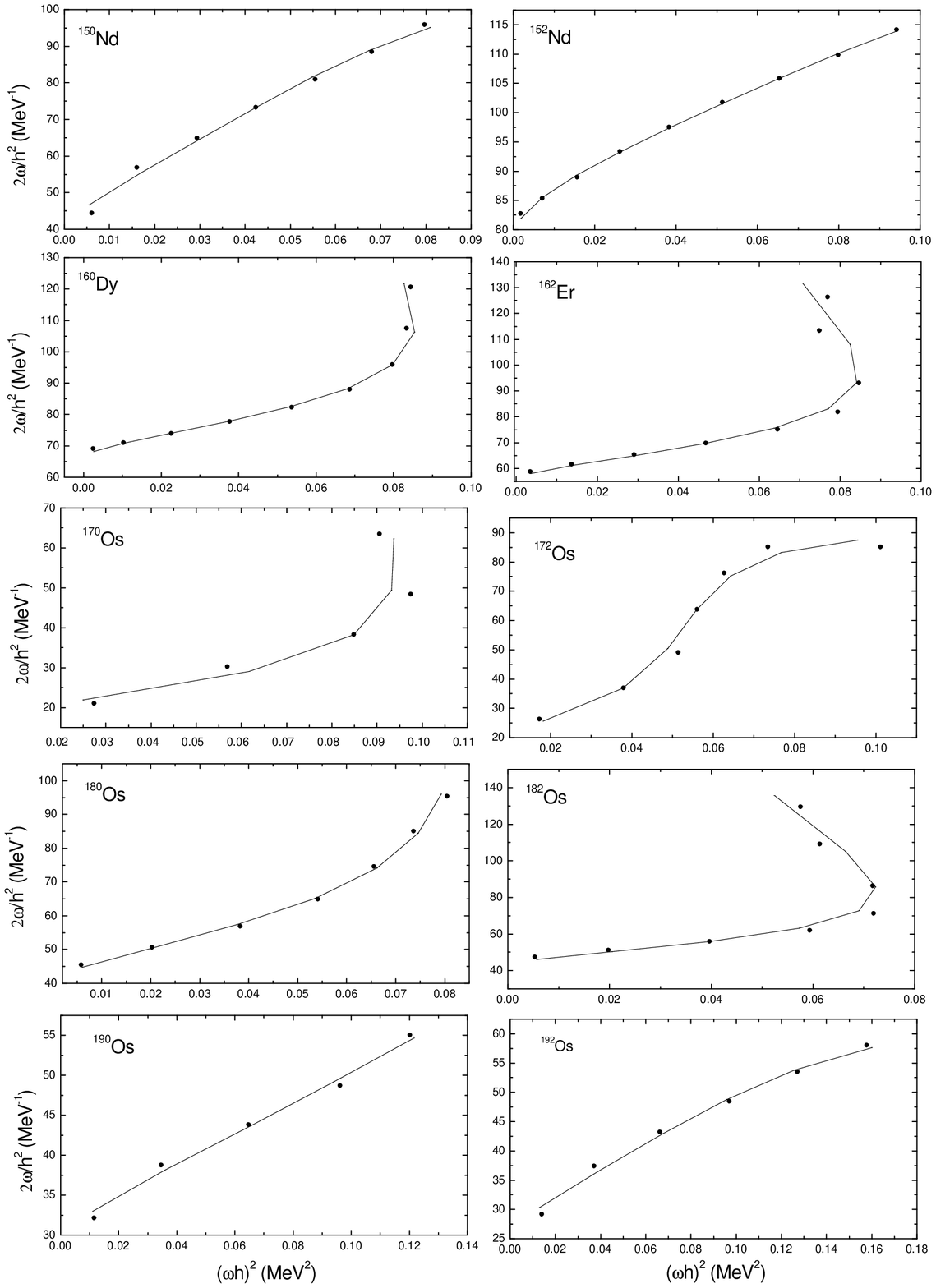}
\caption{The moment of enertia 2$\protect\varphi /\hbar ^{2}$ versus the
square of the rotational frequency ($\hbar \protect\omega )^{2}$ for $%
^{150,152}$Nd, $^{160}$Dy, $^{162}$Er, and $^{170-192}$Os. Dots represent
experimental values.}
\label{Fig:Figure2}
\end{figure}

\begin{figure}[h!]
\centering
\includegraphics*[clip,width=2.7449in, height=3in]{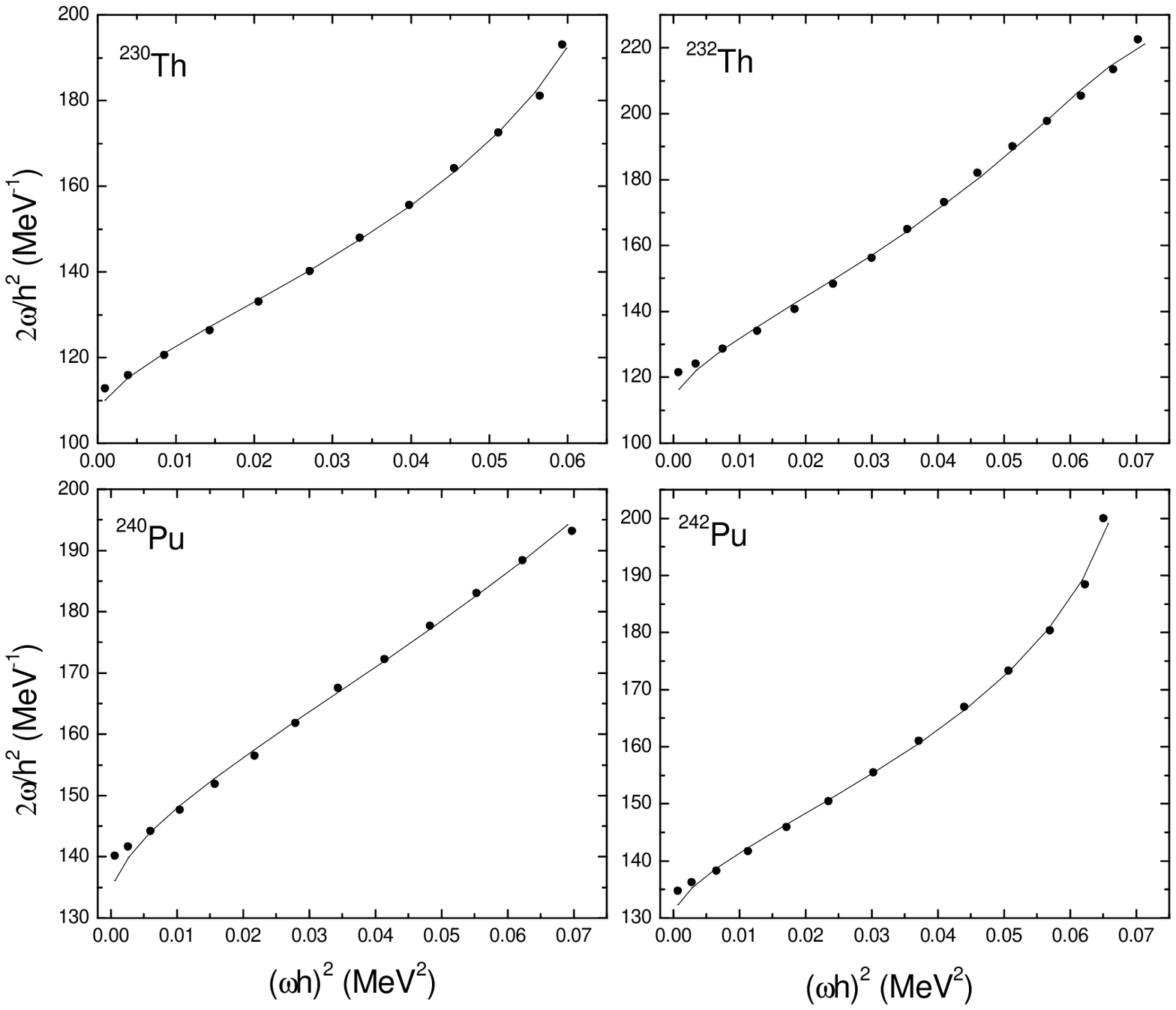}
\caption{The moment of enertia 2$\protect\varphi /\hbar ^{2}$ versus the
square of the rotational frequency ($\hbar \protect\omega )^{2}$ for $%
^{230,232}$Th and $^{240,242}$Pu. Dots represent experimental values}
\label{Fig:Figure3}
\end{figure}

These results show that the present improved version of the exponential
model with paring attenuation is a successful tool in studying ground state
energy levels in slightly deformed and deformed nuclei up to high spin. The
restriction imposed in reference \cite{Sood} in the energy ratio $E\left(
4^{+}\right) /E\left( 2^{+}\right) $ to exceed $3.0$ does not arise here
since the exponent $\nu $ is a free parameter (see last column of table II).
Moreover, our improved model overcomes the inapplicability found in previous
works concerning the description of the forward or down-bending region of
the $\varphi -\omega ^{2}$ plot.

\pagebreak

\end{document}